\documentstyle[prd,aps,floats]{revtex}
\def\spose#1{\hbox to 0pt{#1\hss}}
\def\lta{\mathrel{\spose{\lower 3pt\hbox{$\mathchar"218$}}
     \raise 2.0pt\hbox{$\mathchar"13C$}}}
\def\gta{\mathrel{\spose{\lower 3pt\hbox{$\mathchar"218$}}
    \raise 2.0pt\hbox{$\mathchar"13E$}}}
          \def\z{{\cal Z}}     \def\zet{\zeta}  
\def\ce{c_{\zet}}    \def\wig{\xi}             \def\bb{b}
\def\ms{m_\sigma}          \def\mx{m_{\rm x}}        
\def\mP{m_{_{\rm P}}}       \def\mp{m_{\rm p}}        \def\me{m_{\rm e}}  
       \def\aa{{g^*}}          \def\ls{\lambda_\sigma} 
        \def\I{I}
\def\T{{\cal T}}       
\def\Th{\Theta}    \def\Ta{\Th_\star}   \def\TR{\Th_{_{\rm R}}} 
\def\TE{\Th_{_{\rm E}}}   \def\Tx{\Th_{\rm x}}      \def\Ts{\Th_\sigma}

\begin{document}
\draft

\newcommand{\vp}{\varphi}
\newcommand{\be}{\begin{equation}} \newcommand{\ee}{\end{equation}}
\newcommand{\bea}{\begin{eqnarray}} \newcommand{\eea}{\end{eqnarray}}

\renewcommand{\topfraction}{0.99}
\renewcommand{\bottomfraction}{0.99}
\twocolumn[\hsize\textwidth\columnwidth\hsize\csname 
@twocolumnfalse\endcsname

\title{Prolongation of Friction Dominated Evolution \\
for Superconducting Cosmic Strings}

\author{B. Carter$^1 \,$\cite{maila},  R. Brandenberger$^2 \,$\cite{mailb}, A.-C. Davis$^3 \,$\cite{mailc} and G. Sigl$^1 \,$\cite{maild}}
\bigskip
\address{$^1$ D.A.R.C., Centre National de la Recherche Scientifique,\\
Observatoire de Paris-Meudon, 92 195 Meudon, France;\\
$^2$ Physics Department, Brown University,\\ Providence, R.I. 02912, USA;\\
$^3$ DAMTP, Center for Mathematical Sciences, University of Cambridge,\\
Wilburforce Road, Cambridge, CB3 0WA, U.K.}

\date{18 September, 2000}

\maketitle

\begin{abstract}
This investigation is concerned with cosmological scenarios based on
particle physics theories that give rise to superconducting cosmic strings 
(whose subsequent evolution may produce stable loop configurations known as 
vortons). Cases in which electromagnetic coupling of the string
current is absent or unimportant have been dealt with in previous work.
The purpose of the present work is to provide quantitative estimates
for cases in which electromagnetic interaction with the surrounding plasma 
significantly affects the string dynamics. In particular it will be
shown that the current can become sufficiently strong for the initial
period of friction dominated string motion to be substantially prolonged,
which would entail a reinforcement of the short length scale end of 
the spectrum of the string distribution, with potentially observable 
cosmological implications if the friction dominated scenario lasts 
until the time of plasma recombination.

\end{abstract}

]


\vskip 0.4cm

\section{Introduction}
\label{sec:0}

The original motivation for this work was the investigation of conditions 
under which it might be necessary to revise a previous
analysis~\cite{BCDT96} of the vorton production that is likely to occur
in the framework of various categories of ``superconducting'' cosmic
string forming particle physics theories. This used  a simplified
picture in which the strings themselves arose as vortex
defects of the vacuum formed by spontaneous symmetry breaking
characterised by a mass scale $\mx$ at a corresponding cosmological
temperature 
\be \Tx\approx \mx\ ,
\label{CD0}\ee
after which a (Witten type)
``superconducting'' current is supposed to condense on the string when
the cosmological temperature $\Th$ drops below another mass scale,
$ \ms\lta \mx$ (where for example $\mx$ might be as high as the GUT
scale, of the order of $10^{16}$ GeV, while $\ms$ might be as low as
the electroweak scale, of the order of $10^2$ GeV).

It has already been shown~\cite{CD00} that a revised estimate (according
to which the vorton production is substantially enhanced) is required
for cases in which the string is of the special chiral type~\cite{CP99},
which can occur within the framework of many supersymmetric theories,
but only when the relevant string current is of electromagnetically
uncoupled kind.  The present analysis was motivated by a question of a 
very different kind, concerning the kind of revision that might be 
relevant when the string current is of an electromagnetically coupled kind. 

The preceding analysis~\cite{BCDT96} distinguished between two qualitatively 
different scenarios, depending on whether or not the current condensation 
temperature 
\be \Ts\approx \ms 
\label{CD1}\ee
is sufficiently high to occur in the {\it friction dominated regime} 
that is characterised~\cite{Kibble82} by the condition 
\be \tau\lta t 
\label{CD2}\ee
to the effect that the relevant friction damping time scale $\tau$ say
should be short compared with the cosmological time scale $t\approx
H^{-1}$, where, according to the standard Friedmann formula, the Hubble
expansion rate $H$ is given in terms of the mean cosmological
mass-energy density $\rho$ by $H^2\simeq 8\pi G\rho/3$ where $G$ is
Newton's constant.  Using units such that the speed of light $c$ and
the Dirac-Planck constant $\hbar$ are both set equal to unity, this
cosmological time scale is given during the radiation dominated era
(during which the transitions under consideration would have occurred)
by an expression of the well known form
\be
t\approx {\mP\over\sqrt{\rho}} \ ,
\label{CD3}\ee 
in which the mass energy density of the radiation is given in terms
of the cosmological temperature $\Th$ by
\be \rho\approx\aa\Th^4 
\label{CD4}\ee
where $\aa$ is the effective number of massless degrees of freedom in
the temperature range under consideration.  Note that $\aa\approx 1$ at
low temperatures but that in the range where vorton production is
likely to occur, from electroweak unification through to grand
unification, something more like $\aa \simeq 10^2$ would be a
reasonable estimate.

The previous analysis~\cite{BCDT96} was based on the traditional
supposition~\cite{VV84,V91} that the relevant friction damping time scale 
$\tau$ would have been given in terms of the string tension
\be \T\approx \mx^{\, 2}\
\label{CD5}\ee
 by an expression of the form
\be
\tau\approx{\T \over\beta\,\Th^3} \ .
\label{CD6}\ee
Here $\beta$ is a dimensionless drag coefficient that was assumed to
have an approximately constant value, 
\be
\beta\approx \beta_{\sigma} 
\label{CD7}\ee
depending on the details of the underlying field theory but
typically estimated~\cite{V91} to be of the order of unity,
\be
 \beta_{\sigma} \approx 1 
\label{CD8}\ee
on the assumption that the dominant drag effect would be due to
direct scattering of ambient thermal gas particles by the string core.

It has recently been pointed out by Dimopoulos and Davis~\cite{DD98}
that if the string carried
an electromagnetic current with sufficiently large local (root mean
square) magnitude, $\I$ say, then this standard picture would need to be
revised. The purpose of the present article is to demonstrate that there
are many cases where the current will indeed become sufficiently large for
such a revision to be needed, and to provide quantitative estimates
showing how the traditional picture of string evolution will need to
be modified in such cases. 

It will be shown below that there are indeed many scenarios in which 
electromagnetic currents can engender cosmologically interesting 
consequences~\cite{DD99} by substantially modifying the evolution of the 
main part of the string distribution.  However it would  appear that the 
effect of such modifications will always be  either too little or else 
too late to invalidate the estimates of vorton production~\cite{BCDT96} 
that were cited above. (It is however to be remarked that this confirmation 
of preceeding work on vorton production should not be considered to be 
definitive, since as well as the neglect of electromagnetic drag, which is 
shown here to be generally justifiable, this work~\cite{BCDT96} also 
involved other simplifying assumptions and approximations that still 
need further investigation.)

The outline of this paper is as follows: In the following section we
discuss the new physical effect which leads to the changes in the
cosmological evolution of the vorton density, namely plasma drag. It
follows that if the current $I(t)$ becomes larger than the critical
current $I_c$, the friction dominated regime of the cosmic string
network will endure indefinitely. In Section 3 we derive the 
expression for the {\it wiggle smoothing length $\xi$}, the crucial 
length scale of the string network which determines the density in 
strings. In Section 4 we discuss the initial current on the strings. It 
is in this section that the crucial differences compared to the original
analysis of Dimopoulos and Davis~\cite{DD98} arise. In Section 5 we follow
the evolution of the current during the string network evolution. An 
important parameter which enters the analysis at this point is the {\it 
blueshift factor} ${\cal Z}$. The value of this factor is derived in the 
subsequent section. In Section 7, the results of the two previous sections 
are combined to derive the expression for the current $I(t)$ on the string. 
The case when $I(t)$ remains smaller than $I_c$ is analyzed in Section 8, 
the other case is discussed in Section 9. Our results are summarized in 
Section 10.

A word concerning the notation adopted in this paper: We work in units in
which the both speed of light and Planck's constant are set to 1. Newton's
constant $G$ is not set equal to 1, and therefore the Planck mass $m_p$
often arises in our formulas. Perhaps unconventionally, we denote the
temperature by $\Th$ and reserve the symbol $\T$ for the string tension. 
The cosmological time is denoted by $t$.

\section{Plasma drag on an electrically conducting string}
\label{sec:1}

The reason why the traditional picture will need to be modified in cases
when the string current becomes sufficiently high is the
resulting enhancement of friction
drag exerted on a relativistically moving cosmic string by its surrounding
thermal background.  The effect of the drag contribution that will be 
produced by electromagnetic interaction with the surrounding plasma, 
will be characterised by a friction damping  time scale $\tau$ whose
order of magnitude has been estimated~\cite{DD98} to be given by
\be
\tau\approx{\T \over\I\,\sqrt\rho} \ . 
\label{CD9}\ee
The difference in the friction damping time arises because an 
electromagnetically coupled superconducting string creates a magnetic field 
around the string core. The field is of sufficient strength that it doesn't 
allow the plasma particles to approach the string, creating a magnetocylinder 
around the string, whose radius increases with the string current. This 
results in a much larger scattering cross-section than in the 
non-electromagnetically coupled case and hence a stronger frictional drag.
This plasma friction damping formula (\ref{CD9}) is consistent with the 
preceeding general formula (\ref{CD6}) if, instead of using the constant 
(\ref{CD7}), we take the dimensionless coefficient $\beta$ to have a 
variable value given by the expression
\be \beta\approx {\sqrt{\aa}\I\over\Th}\ ,
\label{CD10}\ee
which will be applicable whenever it gives a value large compared with the 
order of unit value given by (\ref{CD8}). On the other hand in what we shall 
refer to as the low current regime, when (\ref{CD10}) gives a result that 
is small compared with unity, it is the traditional formula, as given 
by (\ref{CD7}), that will be applicable.

Whichever formula applies, i.e. whether or not the coefficient $\beta$
is large compared with unity, it can be seen that it can be used to 
express the ratio of the cosmological time scale $t$ to the damping time scale 
$\tau$ in the form
\be {t\over\tau}\approx {\mP\,\beta\,\Th\over \sqrt{\aa}\,\mx^{\,2}}
\label{CD11a}\ee

In the traditional picture characterised by (\ref{CD7}), the friction
dominated regime (\ref{CD2}), for which the ratio (\ref{CD11a}) is large,
comes to an end when the temperature $\Th$ drops below 
the Kibble limit value $\Ta$ given by 
\be
\Ta \approx\frac{\sqrt{\aa}\, \Tx^{\,2}}{\beta_{\sigma}\,\mP} \ .
\label{CD11b}\ee
However in the plasma damping scenario characterised by (\ref{CD9})
it can be seen from (\ref{CD3}) that one will have
\be 
{t\over\tau}\approx {\mP\, \I\over \mx^{\,2}}\ ,
\label{CD12}\ee
so it follows, as was observed by Dimopoulos and Davis, that if $\I$
remains greater than a certain critical lower limit value,
\be \I \gta I_{\rm c}
\label{CD13} \ee
where~\cite{DD98} 
\be \I_{\rm c}\approx {\mx^{\,2}\over \mP}\ ,
\label{CD14} \ee
then according to (\ref{CD12}) the inequality (\ref{CD2}) will always
be satisfied. Thus if (\ref{CD13}) continued to hold, then the friction
dominated regime would endure indefinitely (at least until the end of
the radiation dominated era, when the foregoing assumptions will cease
to be applicable). On the other hand whenever (\ref{CD13}) is not satisfied,
electromagnetic friction damping will be effectively negligible.

One of the main purposes of the present work is to investigate the
conditions under which the current $\I$ actually can develop an
amplitude exceeding the critical value (\ref{CD14}) above which
electromagnetic drag is important.

\section{Evaluation of the wiggle smoothing length}
\label{sec:2}

Before proceeding to the estimation of the magnitude of the current 
amplitude  $\I$ itself, it will be useful to consider how it influences
the wiggle smoothing length scale $\wig$ (denoted by $R$ in the 
preceding work~\cite{DD98} and by $L_{\rm min}$ in the previous 
analysis~\cite{BCDT96}). In the friction dominated epoch, this scale is 
(roughly) defined as the minimum value for a length scale $L$ for 
which string wiggles of significant amplitude are present, i.e. $\wig$ 
is the length scale below which dissipative ironing out has had time to 
be effective. One of the reasons why this  {\it extrinsic} wiggle smoothing 
length $\wig$ (which should not be confused with the {\it internal} 
electromagnetic smoothing length $\zet$ introduced below) is important is 
that it determines the string energy per unit volume, $\rho_{\rm s}$ say, 
which can be seen by elementary dimensional considerations~\cite{VS94,BCDT96} 
to be given in order of magnitude by an expression of the form
\be \rho_{\rm s}\approx \T\nu\xi^{-2}\ ,
\label{CD28}\ee
where $\nu$ is a dimensionless parameter, that will remain approximately
constant with a magnitude of order of unity so long as resistive damping
is dominant, but that will acquire a much smaller value if and when
radiative damping takes over as the dominant loss mechanism. 

According to the generally accepted picture~\cite{VS94}, during the 
friction dominated regime (which would last at least so long as 
the temperature exceeds the Kibble limit (\ref{CD11b}) even if 
electromagnetic damping were negligible) the string distribution 
will be of Brownian type on all length scales $L$ above the relevant 
wiggle smoothing length $\wig$, which will be given in terms of the 
cosmological time $t$ and the relevant friction damping time scale 
$\tau$ by 
\be \wig\approx (\tau t)^{1/2}\ .
\label{CD27}\ee
In these circumstances the factor $\nu$ in (\ref{CD28}) will have a 
value given roughly by 
\be \nu\approx \nu_{\star}\label{CD30}\ee
where $\nu_\star$ is a constant of the order of unity, and it can be
seen from (\ref{CD3}) and (\ref{CD6}) that we shall have
\be \wig^2\approx {\mx^{\,2}\mP\over\sqrt{\aa}\beta \Th^5}
\label{CD27b}\ee 
where $\beta$ is given by (\ref{CD7}) or (\ref{CD10}) as the case
may be.

In a radiation damping epoch, if there is one, the corresponding
formula for the wiggle smoothing length scale will be given, according 
to the preceding analysis~\cite{BCDT96} (after a transition period 
\footnote{See Figure 1 in \cite{BCDT96} for a sketch of the time 
evolution of $\wig$. During the transition period the motion of the 
strings is relativistic, leading the string separation to keep up with 
the Hubble radius. However, the wiggles on the strings cannot be erased 
until their scale falls below a new scale $\wig$ set by the dominant 
energy loss mechanism.} during which a scaling solution is established)
by an expression of the form
\be \wig\approx\kappa t \label{CD27c}\ ,\ee
while the factor $\nu$ in (\ref{CD28}) will  be given by
\be \nu \approx \nu_\star \kappa^{3/2} \, ,\label{CD30b}\ee
where $\kappa$ is a coefficient that will be given in terms of some
order of unity efficiency factor $\Gamma$ by 
\be \kappa\approx \Gamma \Big({\mx\over\mP}\Big)^2\label{CD31a}\ee
when the dominant radiation loss mechanism is gravitational
(as assumed in the preceeding analysis). Note that in the radiation
dominated epoch the energy density is no longer determined by $\wig$,
but instead by the inter-string separation. 

The formula (\ref{CD31a}) would presumably have to be replaced by
an expression of the analogous form
\be \kappa\approx \Gamma \Big({\I\over\mx}\Big)^2\label{CD31b}\ ,\ee
when the dominant radiation loss mechanism is electromagnetic , i.e.
whenever the latter is larger. It is however to be observed that
(assuming the efficiency factor $\Gamma$ is comparable in the
electromagnetic case to what it is in the gravitational case)
the condition for the electromagnetic radiation damping coefficient
(\ref{CD31b}) to dominate its gravitational analogue (\ref{CD31a})
will be the same as the condition (\ref{CD13}) that is sufficient
to guarantee friction dominance, at least so long as the temperature
is above the Rydberg plasma recombination threshold. This means
that in practice the formula (\ref{CD31b}) will never be needed
before the recent low temperature (less than a few e.V.) matter
dominated epoch. Throughout the preceeding radiation dominated epoch,
which is what we are interested in here, it is the gravitational
radiation damping formula (\ref{CD31a}) that will be relevant if
and when friction damping becomes unimportant.

It is to be remarked that in a competition between radiation damping 
mechanisms the dominant one is that which gives the longest smoothing
length. This contrasts with the situation in a competition between
friction drag damping mechanisms, or between a friction drag mechanism
and a radiating damping mechanism, for which the dominant mechanism will
be that which gives the shortest smoothing length, due to the fact that
the friction not only damps out the short wiggles but also effectively
freezes in, and thus actively preserves, long wiggles.

\section{Estimation of the initial current magnitude}
\label{sec:3}

Whereas the previous analysis~\cite{BCDT96} was based on the
simplification of ignoring the electromagnetic (as opposed to inertial)
effects of the relevant string currents, on the other hand the more recent
work of Dimopoulos and Davis~\cite{DD98} was based on a no less drastic
simplification according to which the relevant string current magnitude
$\I$ was supposed to remain constant, with value comparable to Witten's 
predicted~\cite{W85} upper limit, $\I_{\rm max}\approx \I_{\rm x}$ with
$I_{\rm max}\approx I_{\rm x}$
 \be \I_{\rm x}\approx e \mx
\label{CD15}\ee 
for a bosonic current, or the rather stricter upper limit 
$\I_{\rm max}\approx \I_{\sigma}$ with 
\be\I_{\sigma}\approx e \ms
\label{CD16}\ee 
for a fermionic current, where $e\simeq 1/\sqrt{137}$ is the charge 
coupling constant.

The aim of the present work is to develop a more realistic intermediate 
picture, which not only allows for the effect of the electromagnetic 
coupling of the current but also allows for the dissipative damping of 
its amplitude.

What one expects, on dimensional grounds~\cite{BCDT96}, is that at the 
time of its formation, when the cosmological temperature passes through 
the value (\ref{CD1}) characterised by the relevant carrier mass $\ms$,
the current carrying field will be characterised by random fluctuations
with wavelength
\be \ls\approx \Ts^{-1}\ ,
\label{CD17}\ee
and that it will have a root mean square amplitude given by
$\I\approx \I_{\sigma}$, which means that it will initially be
close to saturation in the fermionic case -- though not in the
bosonic case if the Kibble mass scale $\mx$ characterising the string
itself is substantially larger than the carrier mass scale $\ms$.

It is to be remarked that the ratio of this initial (and in the fermionic 
case maximal) current value $\I_{\sigma}$ to the critical value $\I_{\rm c}$ 
given by (\ref{CD14}), above which drag becomes important, will be given
by
\be {\I_\sigma\over\I_{\rm c}}\approx\sqrt{ {\aa e^2\over\beta_\sigma^{\,2}}}
\Big ({\Ts\over\Ta}\Big)\, ,\label{CD16b}\ee
Since the numerical coefficient $\sqrt{\aa e^2/\beta_\sigma^{\,2}}$ will have 
an order of magnitude comparable with unity at most, this means that 
$\I_{\sigma}$ will fall short of the critical value $\I_{\rm c}$ if $\Ts$
is below the Kibble limit temperature $\Ta$.

\section{Evolution of the current magnitude}
\label{sec:4}

Due to the subsequent contraction of the string
-- which occurs despite the effect of cosmological expansion due
to the various frictional (or at a later stage, radiative)
wiggle damping mechanisms to which it is subjected -- the
comoving wavelength of the initial wavelength will gradually shrink 
to a time dependent value
\be \lambda=\z\ls
\label{CD18}\ee
where (using the same notation as in the previous analysis~\cite{BCDT96})
$\z$ is the relevant time dependent blueshift factor. 

If this blueshifting were the only process going on, the local root mean 
square current amplitude would be amplified to a magnitude
of the order of $\I\approx e\lambda^{-1} \approx \z^{-1}\I_{\sigma} $, 
a value that needs to be distinguished from the mean value, 
$\langle\I\rangle_{\{L\}}$, averaged over a macroscopic length scale $L$ 
say, which will be obtainable (as the result of a random walk process) 
as 
\be\langle\I\rangle_{\{L\}} \approx \z^{-1}\I_{\sigma}(\lambda/L)^{1/2}
\approx e (\lambda L)^{-1/2}\ .
\label{CD19}\ee
It is this -- much smaller -- large scale mean value  $\langle\I\rangle$ 
that would be relevant for estimating the charge on a protovoton loop 
detaching with initial circumference $L$, but on the other hand, as 
Dimopoulos and Davis have emphasised it is the (larger) local root mean 
square current magnitude $\I$ that is relevant for the estimation of the 
frictional drag.  Nevertheless it can not be expected to be quite so large as 
they assumed~\cite{DD98}, because, despite the blue shifting of the relevant 
microscopic fluctuation length scale $\lambda$, the real value of the local 
root mean square current $\I$ will in practice become smaller than the 
preceeding estimate $\I\approx e/\lambda$.

The reason why $\I$ will become small compared with $e\lambda^{-1}$
is that in the meanwhile, while dissipative shrinking characterised
by the blueshift factor $\z$ is taking place, other dissipation mechanisms 
will be damping the amplitude of the current fluctuations, with a 
characteristic time scale that will naturally be shortest for 
fluctuations with the shortest wavelengths. This means that the current
fluctuations will end up by being effectively smoothed out on all
length scales $L$ below some critical smoothing length, $\zeta$ say,
that will increase monotonically with time. Assuming the carrier 
current on the string is conserved, the effect of such smoothing will be 
to reduce the root mean square value $\I$ of the current amplitude from 
the undamped value $e/\lambda$, (where $\lambda$, as given by (\ref{CD18}), 
is the adjusted wavelength of the initial fluctuations) to a magnitude 
identifiable with the average value $\langle\I\rangle_{\{L\}}$ over the 
minimum undamped length scale $L\approx \zeta$. One thus obtains the estimate
\be \I\approx \langle\I\rangle_{\{\zeta\}}\approx { e \over
\sqrt{\lambda \zeta}}\ . \label{CD21}\ee

In a vacuum background the relevant dissipation mechanism would presumably
just be electromagnetic radiation back-reaction, for which a crude
dimensional estimate of the corresponding smoothing length $\zeta$
will be given by an expression of the form
\be \zeta \approx \ce\, t
\label{CD20}\ee
where $\ce$ is an electromagnetic smoothing speed that will be
comparable with the speed of light, i.e. $\ce\approx 1$.
However the regime of interest here is that of the early  universe,
at temperatures large compared with the Rydberg ionisation energy,
for which the relevant thermal background will be a plasma.
In the context of a rather similar physical configuration in such a 
plasma background, it has been remarked by Sigl et al~\cite{Sigl97}
that the main energy loss mechanism will arise from resistance to
ambient electric charge displacements (attributable to Debye type
screening oscillations) for which the relevant damping time scale
$\tau_{\zet}$ will be given in terms of the corresponding wavelength,
$\lambda_{\zet}$ say, of the current fluctuations by an expression
of the form 
\be \tau_{\zet}\approx {\sigma_{\rm e}\lambda_{\zet}^{\, 2}}
 \, ,\label{CD20a}\ee
where $\sigma_{\rm e}$ is the electrical conductivity. A rough
estimate~\cite{Ahonen96} of the effective conductivity in such a 
cosmological plasma is provided by an expression of the form
\be  \sigma_{\rm e}\approx \sqrt{\aa}\Th
\, ,\label{CD20b}\ee
in which it is to be remarked that the factor $\sqrt{\aa}$ is of
the order of ten in most of the regime of interest, but that it drops
to the order of unity after the temperature  $\Th$ falls below that of
electroweak unification. Identifying the cut off $\zeta$ with the
value of the wavelength $\lambda_{\zet}$ for which the decay time scale
$\tau$ becomes comparable with the age $t$ of the universe, it can be
seen to follow from (\ref{CD20a}) that its order of magnitude
will be given by 
\be \zet^2 \approx { t\over\sqrt{\aa}\Th}
\, .\label{CD20c}\ee
This can be seen from  (\ref{CD3}) to be equivalent to
taking the smoothing velocity in (\ref{CD20}) to be given by
\be \ce^{\, 2}\approx {\Th\over \mP}
\, ,\label{CD20d}\ee
so that one finally obtains an expression of the explicit form
\be \zeta\approx\sqrt{2\pi\mP\over\aa\Th^3}
\label{CD20f}\ee 
for the magnitude of the electromagnetic smoothing length itself.

Whichever dissipation mechanism is responsible, it can be seen by
 substituting (\ref{CD20}) into (\ref{CD21}) that the mean squared
current magnitude will be given by an expression of the form
\be \I^2\approx { e^2\sqrt{\aa} \over \z \ce  } 
\Big({\ms\over\mP}\Big)\Th^2 \, ,
\label{CD22}\ee
in which, for a vacuum background, one would simply have $\ce\approx 1$,
whereas in the plasma background of the radiation dominated regime of 
interest here the value of $\ce$ will be given by (\ref{CD20d})
so that one obtains
\be \big({\I\over \I_{\sigma}}\big)^2\approx \sqrt{\aa\ms\over\mP}
\big({\Theta\over\ms}\big)^{3/2}{1\over \z}\, .
\label{CD22a}\ee

The condition for the ordinary friction drag contribution, as 
characterised by the damping time scale (\ref{CD6}) to be dominated by
the electromagnetic drag contribution, as characterised by (\ref{CD9}),
is that the latter time scale should be shorter, or equivalently
that the corresponding dimensionless coefficient (\ref{CD10})
should be large compared to unity. This coefficient will be given by
\be \beta\approx \bb
\sqrt{\ms\over\mP\,\z} \ .
\label{CD23}\ee
where $\bb$ is another positive dimensionless coefficient given by
\be \bb^2={e^2\aa^{3/2}\over\ce}
\label{CD24}\ee
 It can be seen that (contrary to what would
follow~\cite{DD98} if the current were supposed to retain a constant 
amplitude of the order of the fermionic saturation value $\I_{\sigma}$)
the electromagnetic friction damping contribution will remain negligible
unless and until the blue shift factor gets below a limit given
by 
\be \z\sqrt{\Th\over\ms}\lta e^2\aa^{3/2}\sqrt{\ms\over\mP}
\, .\label{CD24a}\ee

\section{Estimation of the blue shift factor}
\label{sec:5}

In order to use the preceding formula to evaluate the root mean square
current amplitude $\I$ as an explicit function of the cosmological
temperature $\Th$ it is necessary to know how the blue shift 
factor $\z$ evolves as a function of the string mass per unit volume, 
$\rho_{\rm s}$ as given by (\ref{CD28}).  If the evolution were an entirely  
continuous process, the factor $\z$  would simply be proportional to
the total comoving string length, 
\be \Sigma\approx {\rho_{\rm s}\over \T \Th^3}\ ,
\label{CD25}\ee 
in a comoving volume characterised by the thermal length scale
$\Th^{-1}$, but (as discussed in the previous work~\cite{BCDT96})
due to the fact that part of the string length will be lost in the 
form of small loops chopped off discontinuously at intersections, 
the blueshift factor will actually be given by an expression of the form 
\be \z\approx \Big({\Sigma\over\Sigma_{\sigma}}\Big)^{1-\varepsilon} \ ,
\label{CD26}\ee
where $\Sigma_{\sigma}$ is the value of the string length $\Sigma$
at time of the current condensation at temperature $\Ts$, and
where the possibility of string loss into small loops is allowed for
by the presence of the dimensionless index $\varepsilon$.
This ``loop formation efficiency factor'' is difficult to estimate 
precisely, but can be expected to lie in the range 
$ 0\leq \varepsilon\lta 1$.

Provided the condition (\ref{CD2}) for friction dominance is satisfied --
i.e. provided the temperature $\Th$ exceeds the Kibble value $\Ta$ given by
(\ref{CD11b}) or the current exceeds the Dimopoulos - Davis limit
(\ref{CD14}) --  
it can be seen from (\ref{CD28}) and (\ref{CD27b}) that we shall have
\be \Sigma\approx {\nu_\star\sqrt{\aa}\beta\Th^2\over\mP\mx^{\,2}}
\, ,\label{CD29a}\ee
which in the electromagnetic drag dominated case can be seen by (\ref{CD10})
to give
\be \Sigma\approx{\nu_\star\aa \I\Th\over\mP\mx^{\,2}}\, .\label{CD29e}\ee
Regardless of which drag mechanism is dominant,
 it follows that if the conditions for friction dominance were also
satisfied  at the time of the current condensation, i.e.
if $\Ts\gta \Ta$, then (independently of the  value of  the dimensionless 
coefficient $\nu_{\star}$)  the blueshift factor will be given by
\be \z\approx\Big({\beta\Th^2\over\beta_\sigma\Ts^2}\Big)^{1-\varepsilon}
\, .\label{CD29}\ee

In a radiation damping epoch, if there is one, the relevant
formula for the wiggle smoothing length scale will be given,
according to the preceding analysis~\cite{BCDT96} by an expression
of the form (\ref{CD27c}), and the factor $\nu$ in (\ref{CD28})) will  
be given by (\ref{CD30b}) in which, as we have seen, the coefficient
$\kappa$ will be given by the gravitational damping formula
(\ref{CD31a}) so that (after the transition period during which a scaling
solution is established) we shall end up with
\be \Sigma\approx{\nu_\star\aa\Th\over \sqrt\Gamma \mP\mx}\, .\label{33}\ee 
If the radiation damping epoch had already begun at the time of
the current condensation, i.e. if $\Ts\lta\Ta$ then it can be seen
that the blue shift factor will be given by
\be \z\approx \Big({\Th\over \Ts}\Big)^{1-\varepsilon}\, .\label{CD33e}\ee
On the other hand for $\Ts\gta\Ta$ a subsequent radiation damping regime,
if any, would be characterised by
\be \z\approx \Big({\sqrt\aa \mx\Th\over\sqrt\Gamma\beta_\sigma
\Ts^{\,2}}\Big)^{1-\varepsilon}\, . \label{CD33f}\ee

A remaining possibility that may be conceived  -- though we shall see that
it will not actually occur in practise -- is that for which the current 
condensation occurred for $\Ts\lta\Ta$, i.e. after the Kibble transition
to a radiation damping regime characterised by (\ref{33}), but for which 
the current subsequently became strong enough to restore 
a regime of friction dominance characterised by (\ref{CD29a}),
in which we would evidently obtain
\be \z\approx\Big({\sqrt\Gamma \I\Th\over \mx\Ts}\Big)^{1-\varepsilon}\, .
\label{34}\ee

\section{Initial low current regime.}
\label{sec:6 }

Just after the current forming transition as the cosmological
temperature $\Th$ drops past the value $\Ts\approx\ms$, it can be seen
that since, by its definition, the blue shift factor will still have
unit magnitude, $\z\approx 1$, at this stage, the immediate effect of
the resistive dissipation will be to reduce the root mean square
current amplitude $\I$ from its initial condensation value
$\I_{\sigma}$ as given by (\ref{CD15}) to a considerably lower value
which according to (\ref{CD22a}) will be given by
\be \big({\I\over \I_{\sigma}}\big)^2 \approx \sqrt{\aa\ms\over\mP}
\ . \label{CD40}\ee
Since our investigation is concerned only with phase
transitions occurring at or below the GUT level we shall have
\be \ms\lta {\mP\over \aa}\ ,\label{CD41}\ee
from which it can immediately be seen that (\ref{CD40}) implies
\be  \I\lta \I_{\sigma} \ , \label{CD42}\ee
and hence that the damped root mean square value $\I$ of the current
amplitude will be below the saturation value $\I_{\rm max}$ not only
in the fermionic case (\ref{CD16}) but also, a fortiori, in the
bosonic case (\ref{CD15}). Furthermore, since, in the usual kinds of
GUT theory (though not of course in Hagedorn type models, which
are not considered here) the order of magnitude of the effective 
number $\aa$ of degrees of freedom in the relevant temperature range
will never exceed the inverse fine structure constant $1/e^2\simeq 137$,
i.e. we shall have
\be \aa e^2\lta 1,\label{CD43}\ee
it can be seen that (\ref{CD40}) also implies
\be \aa \I^2\lta \ms^{\,2}\ , \label{CD44}\ee
which means that the right hand side of (\ref{CD10}) will be small
compared with unity, and therefore that at this stage the electromagnetic
drag contribution will be unimportant. Thus to begin with,
after the condensation of the carrier
particles characterised by the mass scale $\ms$, there will a phase
during which (as assumed in nearly all work prior to that of Davis and 
Dimopoulos~\cite{DD98}) the system will be in what we refer to as a
low current regime, which is characterised by an order of unity drag 
coefficient 
\be \beta\approx 1 \ . \label{CD45}\ee

So long as this low current phase lasts, provided that the temperature
has not yet dropped below the Kibble limit value (\ref{CD11b}), the relevant 
blue shift factor will be given according to (\ref{CD29}) simply by
\be \z\approx\Big({\Th^2\over\ms^2}\Big)^{1-\varepsilon}\ .
\label{CD47}\ee
It therefore follows from (\ref{CD22a}) that the relevant mean squared
current will be given as a rather slowly varying function of the
cosmological temperature $\Th$ by an expression of the form
\be \left({\I\over \I_{\sigma}}\right)^2\approx \sqrt{\aa\ms\over\mP}
\left({\ms\over\Th}\right)^{(1-4\varepsilon)/2}\ .
\label{CD48}\ee

It seems plausible to suppose that the (not yet very well understood)
value of the small dimensionless coefficient $\varepsilon$ would be
less than $^1/_4$ in any regime of friction damping (though it might
well be higher in a regime of radiation reaction damping, during which
a lot of loop creation by string intersections might be expected) but
even if $\varepsilon$ were somewhat larger (in which case (\ref{CD48})
would imply that the absolute value of $\I$ would  actually undergo a
slight decrease as $\Th$ goes down) the right hand side of
(\ref{CD10}), which is proportional to relative value $I/\Th$, would
still increase as $\Th$ goes down. This means that (unlike what was
supposed in nearly all earlier work) the low current regime will not
necessarily last indefinitely, but may be brought to an end when the
cosmological temperature $\Theta$ has fallen to a critical value $\TE$
say, below which the main friction contribution will be of
electromagnetic origin. This will occur when the right hand side of
(\ref{CD10}) reaches the order of unity, so it can be seen that the
relevant electromagnetic transition temperature $\TE$ will be given by
\be \left({\TE\over\ms}\right)^{5-4\varepsilon}\approx
{\aa^3 e^4 \ms\over\mP}\, ,\label{CD49}\ee
provided that this value of $\TE$ is still above the Kibble limit $\Ts$
given by (\ref{CD11b}). This necessary condition for a transition to
an electromagnetic friction drag regime can be seen to be
expressible as
\be \Big({\Ts\over \Ta}\Big)^{5-4\varepsilon}\gta {\mP\over\aa^3 e^4\ms}
\, ,\label {CD49e}\ee
or equivalently
\be\Big({\sqrt\aa\mx\over\beta_\sigma\mP}\Big)^{2-2\varepsilon}\lta
\sqrt{\beta_\sigma} e^2 (\aa)^{5/4}\Big({\ms\over\mx}\Big)^{3-2\varepsilon}
\, .\label{CD49f}\ee
This condition only excludes the cases of extreme disparity for which
the strings formation energy scale $\mx$ is comparable with the GUT
level but the current condensation energy scale $\ms$ is way down nearer
the electroweak level. The condition (\ref{CD49f}) for the occurrence
of an electromagnetic friction dominated regime will be satisfied in most
other kinds of scenario, in which either both the string forming and
the current forming phase transitions occur near the GUT level or else
they both occur at much lower levels.

\section{Final low current regime}
\label{sec:7}

If the condition (\ref{CD49e}) is not satisfied, then when the temperature
has fallen to the Kibble limit value $\Ts$ given by (\ref{CD11b}) it will
enter a radiation damping regime in which the blue shift factor $\z$ will
no longer be governed by (\ref{CD47}). If, as supposed above, the 
current condensation phase transition had already occurred at a temperature
$\Ts$ higher than the Kibble limit value $\Ta$, then $\z$ will be governed
by (\ref{CD33f}) so by (\ref{CD22a}) we shall obtain
\be \big({\I\over \I_{\sigma}}\big)^2\approx \sqrt{\aa\ms\over\mP}
\Big({\sqrt\Gamma\beta_\sigma\ms\over\sqrt\aa \mx}\Big)^{1-\varepsilon}
\big({\Th\over\ms}\big)^{(1+2\varepsilon)/2}\, .
\label{CD50}\ee
This evidently implies that the current amplitude $\I$ will continue to 
decrease as the temperature goes down, so there will be no possibility of 
it building up again to a sufficiently high value for friction drag to 
become important again. The implication of this is that this final low 
current radiation drag dominated regime will last indefinitely.

A similar conclusion, namely that electromagnetic drag will never become
important, will be obtained a fortiori in the case for which the system
is already in the radiation damping regime below the Kibble limit $\Ta$
when the phase transition at the temperature $\Ts$ occurs. In this case
the initial low current stage described in the preceeding section will
be skipped, and the system will pass directly to a final low current regime
in which, instead of by (\ref{CD33f}), the blueshift factor will be governed
by (\ref{CD33e}), so that instead of (\ref{CD50}) the equation governing
the decay of the current amplitude will have the simpler form
\be \big({\I\over \I_{\sigma}}\big)^2\approx \sqrt{\aa\ms\over\mP}
\big({\Th\over\ms}\big)^{(1+2\varepsilon)/2}\, .
\label{CD50b}\ee

\section{Current dominated regime}
\label{sec:8}

The main innovation in the present work is the investigation of cases for 
which the condition (\ref{CD49e}) is satisfied, i.e. for which we have 
\be\TE\gta\Ta\, ,\label{CD51b}\ee
so that after the transition past the threshold (\ref{CD49}) when the 
temperature becomes low enough to satisfy
\be \Th\leq \TE\, ,\label{CD51}\ee
there will occur what we shall refer to as a current dominated regime,
meaning a regime in which the current amplitude $\I$ is large enough for
electromagnetic friction drag to provide the main dissipation mechanism
acting on the cosmic string distribution. A noteworthy feature of such 
a regime is that the ratio of the string mass density $\rho_{\rm s}$ as 
given by (\ref{CD28}) \footnote{Note that in this case there is never 
a radiation-dominated epoch so that the use of (\ref{CD28}) is justified.}
to the total mass density $\rho$ as given by (\ref{CD4}) will be 
expressible, using  (\ref{CD10}) and (\ref{CD27b}), by
\be {\rho_{\rm s}\over\rho}\approx \nu_\star {\I\over \mP}\, ,
\label{CD51e}\ee
(in which it is to be recalled that $\nu_\star$ is just a constant of
the order of unity). Such a regime can only last so long as $\I$
remains above the Dimopoulos-Davis critical value $\I_{\rm c}$ given
by (\ref{CD13}),  a condition which can be seen from (\ref{CD51e}) to be
expressible as
\be {\rho_{\rm s}\over\rho}\gta \nu_\star\Big({\mx\over\mP}\Big)^2\, .
\label{CD51f}\ee
This condition ensures the preservation of wiggle structure on length scales
$L$ down to a minimum smoothing value $\wig$ given according to 
(\ref{CD12}) and (\ref{CD27}) by
\be \wig^2\approx {\I_{\rm c}\over I} t^2\, ,\label{CD51g}\ee
that satisfies the condition of remaining short
\be \xi\lta t\, ,\label{CD51h}\ee
compared with the cosmological length scale $t$.

Under these conditions, the comoving string length $\Sigma$ will be given
by (\ref{CD29e}) so by (\ref{CD26}) the redshift factor will be given by
\be \z\approx \Big({\sqrt{\aa} e\I\Th\over \beta_\sigma \I_\sigma\Ts}
\Big)^{1-\varepsilon} \label{CD52}\, .\ee
However, unlike the corresponding formula (\ref{CD47}) for the preceding
low current regime, the expression (\ref{CD52}) does not, by itself,
specify the explicit temperature dependence of  $\z$, because
it also involves the current amplitude $\I$, whose evolution,
according to  (\ref{CD22a}) is also $\z$ dependent. To obtain the explicit
temperature dependence of these quantities, we start by eliminating 
$\z$ between 
(\ref{CD22a}) and (\ref{CD52}) so as to obtain the relation
\be \left({\I\over \I_{\sigma}}\right)^{3-\varepsilon}\approx 
\Big({\beta_\sigma\over\sqrt{\aa}e}\Big)^{1-\varepsilon}\sqrt{\aa\ms\over\mP}
\left({\Th\over\ms}\right)^{(1+2\varepsilon)/2}\, .
\label{CD53}\ee
The corresponding value for $\z$ will be given by
\be \z^{3-\varepsilon}\approx \left({\aa e^2\over\beta_\sigma^2}
\sqrt{\aa\ms\over\mP} \Big({\Th\over \ms}\big)^{7/2} 
\right)^{1-\varepsilon}\ .\label{CD54}\ee

Since one would expect the loop formation efficiency coefficient $\varepsilon$
to be small in a friction dominated regime, the preceeding formula implies
that the temperature dependence of $\I$ will be rather weak,
but nevertheless it can be seen that in the long run the current amplitude
$\I$ will decrease as the temperature goes down. This means that there
will be no chance of reaching a current saturation regime 
such as was originally envisaged by Dimopoulos and Davis~\cite{DD98}.
However it does not exclude the likelihood that the friction dominated regime 
may be considerably prolonged.

In the very long run of course, the friction dominated regime  must
ultimately be terminated by a transition to the usual kind of radiation 
damping dominated regime, but
if the relevant mass scales are relatively low (compared with the G.U.T,
level) this final transition might not occur until by passage through
the plasma recombination when the temperature reaches the Rydberg level,
\be \TR\approx e^4\me\, ,\label{CD55}\ee
(where $\me$ is the electron mass) but otherwise it will occur at a higher 
temperature when $\I$ gets down to value $\I_{\rm c}$ given by (\ref{CD14}). 
It can be seen from (\ref{CD53}) that this will occur at a 
critical temperature $\Th_{\rm c}$ say given by
\be\Big({\Th_{\rm c}\over\ms}\Big)^{(1+2\varepsilon)/2}\approx
\Big({\sqrt{\aa}\over\beta_\sigma}\Big)^{1-\varepsilon}\sqrt{\mP\over \aa e^4 
\ms} \Big({\mx^{\,2}\over \mP\ms}\Big)^{3-\varepsilon}\, .\label{CD56}\ee
After passing below this critical temperature $T_{\rm c}$, the universe will
find itself in a final low current radiation dominated state of the kind
described in the preceeding section, with the current magnitude governed by
(\ref{CD50}).

\section{Conclusions}
\label{sec:10}

The preceeding analysis shows that the occurrence of an electromagnetic 
friction dominated regime will not have any effect on the vorton production 
estimates whose verification provided the original motivation for this work. 
\footnote{A similar negative conclusion has been reached in unpublished work by
Dimopoulos~\cite{D97}} To understand this result, it is important to 
recall from~\cite{BCDT96} that in the friction-dominated case the vorton 
density is dominated by those produced at the onset of the friction-dominated 
period. From our analysis it has transpired that such an electromagnetic 
friction dominated regime only occurs if the electromagnetic current 
condensation $\Ts$ is higher than the Kibble limit temperature $\Ta$, 
and in that case (though not in general) the dominant vorton production 
mechanism~\cite{BCDT96}  operates immediately and is unaffected by the 
subsequent evolution (friction dominated or otherwise) of the main part of 
the  string distribution.

The fact that it does not affect our present (provisional) understanding of 
vorton formation does not exclude the possibility that the effect of 
electromagnetic friction drag on strings may be cosmologically important. 
In particular -- while the preceeding analysis implies that its occurrence
may be more delicately parameter dependent than was suggested in
when it was originally envisaged by Dimopoulos and Davis~\cite{DD98} --
the possibility evoked in the preceeding section that the friction drag 
dominated epoch may continue until the stage of plasma recombination has 
potentially interesting observational implications in view of the fact that 
this ``last scattering'' period is directly accessible as the source of 
the cosmic thermal background radiation. This will occur if
\be \Th_{\rm c}\lta\TR\, ,\label{CD60}\ee
where $\Th_{\rm c}$ is given by (\ref{CD56}) and $\TR$ by (\ref{CD55}),
or to be more explicit if
\be \Big({\mx\over\mP}\Big)^3\Big({\mx\over\ms}\Big)^{3-2\varepsilon} \lta 
{e^2\beta_\sigma\Big({\sqrt{\aa}\over\beta_\sigma}\Big)^\varepsilon}
\Big({e^4\me\over\mP}\Big)^{(1+2\varepsilon)/2}\, .\label{CD61}\ee
Despite the small value of the Rydberg to Planck energy ratio
$e^4\me/\mP\approx 10^{-26}$ it can be seen that (because of the high
powers of the factors on the left hand side) in cases for which
$\ms\approx\mx$ this condition can easily be satisfied for values
of the string formation energy scale $\mx$ that can be very large
compared with the electroweak level $\ms/\mP\approx 10^{-16}$,
though not if it is too near the GUT level $\ms/\mp\approx 10^{-3}$ 
In such a case the spectrum of the string distribution at the time of 
electromagnetic decoupling would be characterised by the preservation of 
a higher proportion of short wavelength structure than in the traditional 
radiation dominated scenario, and it can be seen from the inequality 
(\ref{CD51f}) that the corresponding string mass density fraction will 
be higher than the traditionally predicted value 
$\rho_{\rm s}/\rho \approx (\mx/\mP)^2$.

\bigskip
{\bf Acknowledgements}
\medskip

Two of us (BC and RB) wish to thank W. Unruh for hospitality at U.B.C., 
Vancouver, where some of this work was carried out. This work is supported
in part by the U.S. DOE under Contract DE-FG0291ER40688, Task A (RB),
by the UK PPARC (ACD) and in part by an ESF network (BC and ACD).

\end{document}